\documentclass[prb,amsmath,amssymb,twocolumn,superscriptaddress,floatfix]{revtex4}
\usepackage{graphicx}
\usepackage{dcolumn}
\usepackage{bm}
\usepackage{subfigure}
\usepackage{amsmath}
\usepackage{feynmf}
\usepackage{hyperref}

\newcommand{\ra}{\rightarrow}

\newcommand{\bs}{\boldsymbol}
\newcommand{\ep}{\varepsilon}

\newcommand{\SRO}{Sr$_2$RuO$_4$}

\usepackage{times}

\begin{document}
\title{Interband interference effects at the edge of a multiband chiral $p$-wave superconductor}
\author{Jia-Long Zhang}
\address{School of Physics, Sun Yat-sen University, Guangzhou, 510275, China}
\author{Wen Huang}
\email{huangw001@mail.tsinghua.edu.cn}
\address{Institute for Advanced Study, Tsinghua University, Beijing, 100084, China}
\author{Manfred Sigrist}
\address{Institut f\"ur Theoretische Physik, ETH-Z\"urich, CH-8093 Z\"urich, Switzerland}
\author{Dao-Xin Yao}
\email{yaodaox@mail.sysu.edu.cn}
\address{School of Physics, Sun Yat-sen University, Guangzhou, 510275, China}
\date{\today}

\begin{abstract}
Chiral superconductors support chiral edge modes and potentially spontaneous edge currents at their boundaries. Motivated by the putative multiband chiral $p$-wave superconductor \SRO, we study the influence of the interference between different bands at the edges, which may appear in the presence of moderate edge disorder or in edge tunneling measurements. We show that interband interference can strongly modify the measurable quantities at the edges when the order parameter exhibits phase difference between the bands. This is illustrated by investigating the edge dispersion and the edge current distribution in the presence of interband mixing, as well as the conductance at a tunneling junction. The results are discussed in connection with the putative chiral $p$-wave superconductor \SRO. In passing, we also discuss similar interference effects in multiband models with other pairing symmetries.
\end{abstract}

\maketitle
\section{Introduction}
\label{sec:Intro}
The problem of the edge current and orbital angular momentum in chiral superfluids has long been studied in the context of the $A$-phase of $^3$He -- a chiral $p$-wave superfluid \cite{Wolfle:77,Leggett:75,Volovik:03}. The interest was renewed after the discovery of superconducting \SRO \cite{Maeno:01,Mackenzie:03,Kallin:09} which shows indications of the same pairing. Such a state is characterized by a nonvanishing topological invariant -- the Chern number \cite{Volovik:03}. Accordingly, the boundaries of these systems hosts chiral Majorana fermions\cite{Volovik:03}, which are expected to carry chiral edge current.


With regard to \SRO, although tunneling measurements have shown evidence of subgap edge states \cite{Laube:00,Kashiwaya:11,Ying:12}, the predicted spontaneous edge current\cite{Matsumoto:99,Furusaki:01} remains elusive \cite{Kirtley:07,Hicks:10,Curran:14}. This poses a severe challenge to the chiral $p$-wave interpretation. Recently, there have been numerous theoretical attempts to address this issue. In particular, due to $U(1)$ symmetry breaking the edge current is not expected to be topologically protected \cite{Huang:15,Tada:16}, hence chiral $p$-wave models with anisotropic gap structure, be it single band or multiband, may generate substantially smaller current compared with that of a simple isotropic chiral $p$-wave model \cite{Ashby:09,Imai1213,Bouhon:14,Lederer:14,Scaffidi:15,Huang:15}. Additional strong suppression is possible in the presence of strong surface disorder \cite{Lederer:14,Scaffidi:15}. Nevertheless, given the signatures of chiral $p$-wave pairing \cite{Maeno:12,Kallin:12,Kallin:16,Mackenzie:17} and the strict experimental upper bound \cite{Kirtley:07,Hicks:10,Curran:14} placed on the magnitude of the edge current (at least three orders of magnitude smaller than the predicted value in Ref \onlinecite{Matsumoto:99}), it is unclear whether the above theories have provided satisfactory explanations for the experimental null results. Interestingly, the edge current may vanish in non-$p$-wave chiral superconductors \cite{Huang:14,Tada:15,Ojanen:16,Suzuki:16}, although these states are less likely for \SRO.

One complexity in \SRO~is its multiband nature. There are three bands crossing the Fermi level \cite{Damascelli:00,Bergemann:00}. In spite of a great deal of studies, questions remain regarding its exact superconducting gap structure on the bands. Nevertheless, there have been numerous examples of multiband or multi-orbital effects giving rise to intriguing physics, such as the collective phase fluctuations between the bands (Leggett mode)\cite{Leggett:66,Huang:16}, a time-reversal symmetry breaking pairing owing to a complex phase configuration on the bands \cite{Agterberg:99}, distinct Josephson effects \cite{Kawai:17}, and a novel type-1.5 behavior indicative of distinct thermodynamic length scales on the different bands \cite{Garaud:12}. In addition, it was shown that the behavior of the local density of states and the order parameter at the boundary can be rather different for models whose band order parameters are in-phase and out-of-phase, even in $s$-wave systems \cite{Tsai:09,Golubov:09,Ghaemi:09,Bobkov:11,Burmistrova:13}. Notably, some of the above examples involve the relative phase degree of freedom peculiar to multiband models. It is thus of considerable interest to also explore how this particular aspect may affect the physics at the edges of multiband chiral superconductors.

In this work, we focus on the interference effects in multiband models with different phase configurations: finite ($\pi$, nontrivial) and vanishing (trivial) phase differences between the multiple band order parameters. We illustrate these with two examples. In the first, we consider the consequence of interference in the form of edge interband mixing. The mixing may be introduced by, e.g. even moderate edge disorder, as will be elaborated later. In certain cases, it introduces hybridization between the quasiparticle states, thereby holding promise for altering the edge current -- of which the edge modes are known to constitute an appreciable portion\cite{Furusaki:01,Stone:04,Sauls:11}. In the second, we study the differential tunneling conductance at a normal metal-superconductor (NS) junction. The models we formulate for these two examples are similar in spirit and both will demonstrate qualitatively distinct interband effects for different phase configurations. In particular, for the nontrivial configuration the interband interference can lead to substantial variations in the physical observables at the boundaries. As an aside, the conclusions also generalize to models with other pairing symmetries, which can also be implicated from an earlier study \cite{Golubov:09}. In particular, the interference in the case the of nontrivial phase configuration can lead to ingap edge states not supported in the corresponding single-band models.

The rest of this paper is organized as follows. In Sec \ref{sec:BdG} we use a numerical Bogoliubov de-Gennes (BdG) calculation to study the effect of edge interband mixing in a two-band chiral $p$-wave superconductor. The results are discussed on the basis of semiclassical and phenomenological theories. We then calculate in Sec \ref{sec:BTK} the tunneling conductance of a two-band chiral superconductor at a NS junction, adopting a semiclassical Blonder-Tinkham-Klapwijk (BTK) approach. Finally, we close with a brief summary in Sec \ref{sec:summary}.



\section{numerical BdG}
\label{sec:BdG}
For simplicity, we focus on spinless two-band chiral $p$-wave models on a two-dimensional square lattice. The results thus obtained can be generalized to other multiband models. We consider the following BdG Hamiltonian,
\begin{equation}
H= H_1 + H_2 + H_{12} \,,
\label{eq:H0}
\end{equation}
where $H_{1 (2)}$ and $H_{12}$ represent the Hamiltonian of band-$1(2)$ and the mixing between the two-bands, respectively. More specifically, for the two bands,
\begin{equation}
H_l =\sum_{\bs k} \left[ \ep_{l,{\bs k}} c^\dagger_{l,{\bs k}} c_{l,{\bs k}} + \Delta_{l,{\bs k}} c^\dagger_{l,{\bs k}} c^\dagger_{l,-{\bs k}}  + \Delta_{l,{\bs k}}^\ast c_{l,-{\bs k}} c_{l,{\bs k}} \right]  \,,
\end{equation}
where again the subscript $l=1,2$ indicates band indices. The band dispersions and gap functions assume the forms $\ep_{l,{\bs k}} = -2t_{l} (\cos k_x + \cos k_y) - 4t^\prime_{l} \cos k_x \cos k_y - \mu_{l}$ and $\Delta_{l,\bs k} = \Delta_{lx} f_{l,\bs k}+ \Delta_{ly}g_{l,\bs k}$. Here $\Delta_{lx}$ and $\Delta_{ly}$ are the two complex components of the superconducting order parameter, and $f_{l,\bs k}$ and $g_{l,\bs k}$ are form factors characteristic of the Cooper pairing with $p_x$ and $p_y$-symmetry. For the chiral pairing on each band, the two components assume a phase difference of $\pi/2$ or $-\pi/2$. We concentrate however on the effects of the overall phase difference between the two band gaps $\Delta_{1,\bs k}$ and $\Delta_{2,\bs k}$, which can be either zero or $\pi$. The latter can be realized if there is repulsive interband Cooper pair scattering.

The actual calculation is performed in a cylinder geometry with open boundaries in the $x$-direction. Physical quantities such as the edge current and the order parameter are determined self-consistently following previous works \cite{Imai1213,Lederer:14}. The interband mixing (\ref{eq:H0}), assumed to be induced by surface disorder, is given by,
\begin{equation}
H_{12} = \sum^N_{i=N^\prime}( t_{12} c^\dagger_{1,i}c_{2,i} +H.c.) \,.
\label{eq:mixing}
\end{equation}
where $i$ is the site index in the $x$-direction. It is nonvanishing only at one of the edges, between sites $i=N^\prime$ and $i=N$. We assume that the mixing is restricted to a small region around the edge much narrower than the coherence length of the two bands. For simplicity, through out this work we take $N^\prime=N$, i.e., finite mixing only at the end site. Note this represents relatively weak surface disorder potentially relevant to samples/devices prepared with high quality but with lattice distortion at the edges (such as the rotation of RuO$_6$ octahedra in the case of \SRO \cite{Matzdorf:00,Veenstra:13}). This is unlike the metallic surface condition, i.e., extremely strong disorder, assumed in other studies \cite{Lederer:14,Scaffidi:15}.

\subsection{BdG results}
\label{sec:Results}
\begin{figure}
\includegraphics[width=8.5cm]{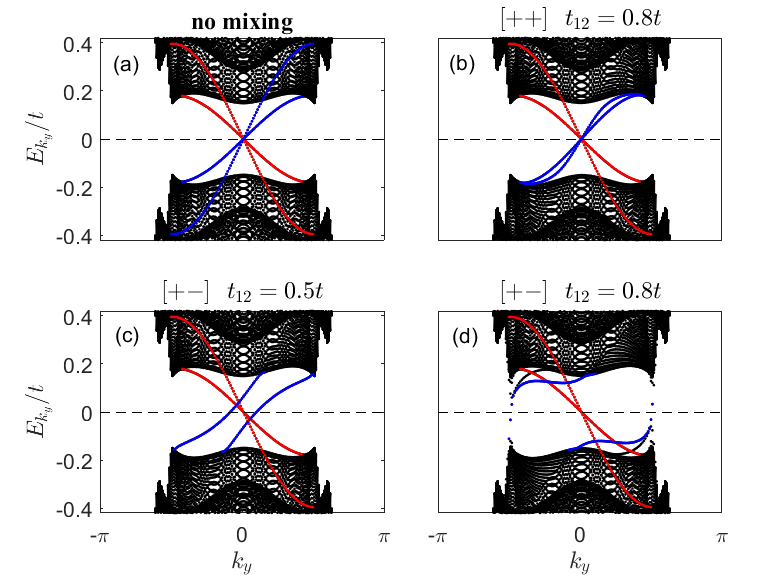}
\caption{(Color online) Low energy spectra of two-band chiral $p$-wave models obtained in self-consistent BdG calculations. In each plot, the edge spectra shown in red (blue) correspond to edge modes near $i=1$ ($i=N$). These calculations were performed on an $N=200$ geometry and assume the following parameters: $t_{1/2}=t=1$, $t^\prime_1=0.375t$, $t^\prime_{2}=0.2t$, $\mu_1=-t$, $\mu_2=-2t$. The chiral $p$-wave pairing takes the form of $\Delta_{l,\bs k}=\Delta_{l}(\sin k_x + i \sin k_y)$ with $|\Delta_{1}|=0.3t$, $|\Delta_{2}|=0.15t$. The $\pm$ symbols in the square brakets denote the phase configuration of the order parameters on the two bands: $[+-]$ and $[++]$ stand for $\text{sgn}[\Delta_{1}]=-\text{sgn}[\Delta_2]$ and $\text{sgn}[\Delta_{1}]=\text{sgn}[\Delta_2]$, respectively. The strength of interband mixing $t_{12}$ at $i=N$ is shown above the plots. }
\label{fig:disp1}
\end{figure}

We base the majority of our discussions on the $T=0$ calculations performed near the continuum limit -- low filling fraction and roughly isotropic gaps, using a set of parameters as described in Fig \ref{fig:disp1}. In the absence of mixing, two chiral edge modes, each associated with one of the two bands,  emerge at each of the two boundaries, as illustrated in Fig. \ref{fig:disp1} (a). The rest of Fig. \ref{fig:disp1} shows the low energy dispersion for models with interband mixing on one of the edges. At finite $t_{12}$, there is a striking distinction between models with different phase configurations on the two bands. In the case of the $[+-]$ configuration, i.e. $\text{sgn}[\Delta_1]=-\text{sgn}[\Delta_2]$, the edge modes around $k_y=0$ readily splits upon introduction of finite $t_{12}$. For sufficiently large $t_{12}$, a large number of the edge states pile up below the continuum edge associated with the smaller gap [Fig \ref{fig:disp1} (d)]. By contrast, for the $[++]$ configuration, i.e. $\text{sgn}[\Delta_{1}]=\text{sgn}[\Delta_2]$, the zero-crossing at $k_y=0$ is nearly unaffected by interband mixing [Fig \ref{fig:disp1} (b)]. Some of the high-energy edge modes originally associated with the larger band gap are pushed down below the continuum edge of the smaller gap [Fig \ref{fig:disp1} (b)]. Overall, there appears to be no spectral flow \cite{Huang:15,Volovik:95,Stone:87} in this scenario, i.e. the occupancy of the states remains largely intact. It is worth stressing that, although the shape of the edge spectrum depends sensitively on the detailed form of the interband mixing used (such as longer-range intersite mixing), the aforementioned general behavior is robust. Some of these qualitative features will also appear in the BTK calculations in Sec \ref{sec:BTK} and will be explained in Sec \ref{subsec:Semiclassical} from a semiclassical perspective.

\begin{figure}
\includegraphics[width=9cm]{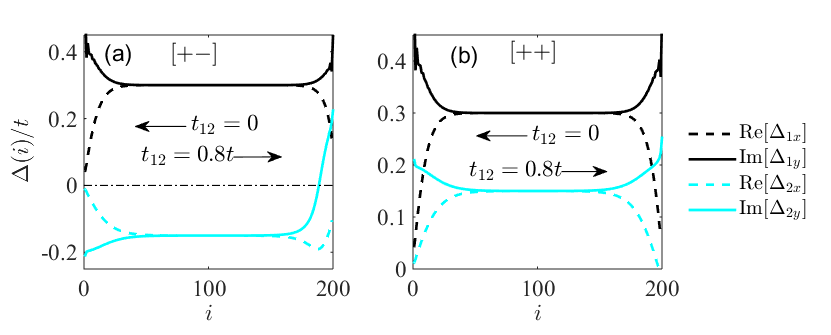}
\caption{(Color online) Self-consistent order parameter profile for models with: (a) $[+-]$ and (b) $[++]$ phase configurations. The interband mixing, $t_{12}=0.8t$, is present only at the right edge. The corresponding low energy dispersion is presented in Figs.\ref{fig:disp1}(d) and \ref{fig:disp1}(b). The $\Delta_x$ components are plotted by dashed lines and the $\Delta_y$ components are shown by solid lines, while the colors designate band $1$ (black) and band $2$ (cyan).}
\label{fig:op}
\end{figure}

The response of the order parameters similarly exhibits a dichotomy between the two phase configurations. In the $[+-]$ model with sizable $t_{12}$, the interference-induced variation is significant and extends over a coherence length of the corresponding bands ($\sim v_{lF}/|\Delta_{l}|$ where $v_{lF}$ is the average $l$-band Fermi velocity). For example, in Fig. \ref{fig:op} (a) one of the $\Delta_y$-components changes sign, in strong contrast to its behavior at the other edge, where $t_{12}=0$. In comparison, in the $[++]$ model no noticeable change is seen beyond the narrow region of mixing. Section \ref{subsec:EFT} presents a phenomenological interpretation of these results.

\begin{figure}
\includegraphics[width=7cm]{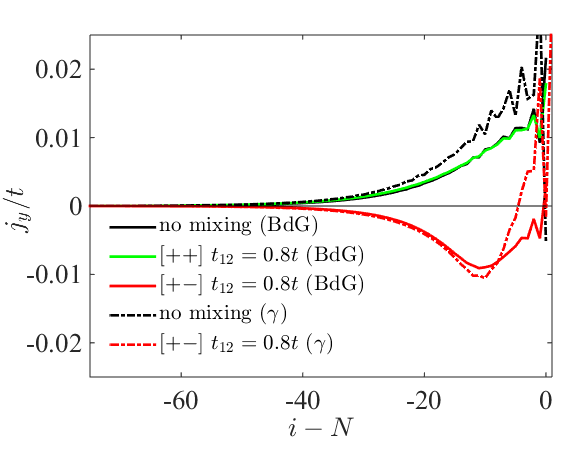}
\caption{(Color online) Edge current distribution corresponding the two-band BdG calculations in Fig.\ref{fig:disp1}. Solid curves (BdG) represent the exact results, while dashed curves ($\gamma$) show estimates based on (\ref{eq:GL}) with $\gamma_1=0.134$ and $\gamma_2=0.103$ evaluated based on the corresponding expression in the main text. Note that in practice the derivative in (\ref{eq:GL}) is replaced by the {\it Euler} approximation. The data without interband mixing ars shown by a black solid line, which is the same for both order parameter phase configurations. We have set $e$ and $\hbar$ to unity, hence $j_y$ is displayed in units of $t$. }
\label{fig:curr}
\end{figure}

The drastic redistribution in the $[+-]$ model of the chiral edge dispersion in momentum and energy space incurs corresponding changes in the edge current (Fig. \ref{fig:curr}). This is easy to understand as the edge modes carry considerable current \cite{Furusaki:01,Stone:04,Sauls:11} (see Sec. \ref{subsec:EFT} for a complementary explanation based on the order parameter variations). By contrast, since there is no apparent spectral flow, the current in the $[++]$ model hardly varies at finite mixing \cite{Huang:15}. What is most striking is the inverted current flow in the $[+-]$ model in the presence of noticeable interband mixing, given that our model employs rather moderate edge disorder (manifest in the fact that the extension of interband mixing is much narrower than the coherence length and that the $\Delta_y$-components survive at the edge in Fig.\ref{fig:op}). Previous one-band studies showed that current inversion is possible by varying the gap structure \cite{Huang:15} and edge orientation in lattice models \cite{Bouhon:14}. Our study therefore demonstrates an additional mechanism, interband interference, for the current to deviate considerably from what is expected of an ideal edge. Similar observations are made for anisotropic multiband chiral $p$-wave models (not shown), which can potentially make contact with multiband models of \SRO.


\subsection{Semiclassical perspective}
\label{subsec:Semiclassical}
We now discuss the interference of the two bands on the quasiparticle level within a first order perturbative theory. The fact that strong splitting of the chiral edge modes occurs at $k_y=0$ for one configuration of $\text{sgn}[\Delta_{1/2}]$ but not for the other is suggestive of some symmetry-related properties of the unhybridized wavefunctions, which are dictated by the phases of the band order parameters. The edge modes are solutions to the BdG equation linearized about the Fermi wavevector. Due to the translational invariance along $y$, each quasiparticle state can be labeled by a momentum $ k_y$. In the Nambu spinor basis $(c_{l,k_y}, c^\dagger_{l,-k_y})^T$, the wavefunction of an edge state in a half-infinite plane at $x\leq 0$ reads,
\begin{equation}
\phi_{l,k_y}(x) \propto \frac{1}{\sqrt{2}}\begin{pmatrix}
\text{sgn}[\Delta_l] \\
-i
\end{pmatrix} \sin(|k_x| x)e^{ik_y y} e^{\lambda x} \,,\,\,\,\, (x\leq 0)\,,
\label{eq:wavefunction}
\end{equation}
where $k_x$ and $k_y$ are components of $\bs k_F$ and $\lambda= |\Delta_{lx0}| f_{l,\bs k}/v_{lx}$. The energy of this mode is given by $E^\text{edge}_{l,k_y} = -\Delta_{ly0}g_{l,\bs k}$. We see that the spinor in (\ref{eq:wavefunction}) is related to the phase of the order parameter and has equal amplitudes in electron and hole components. For an isolated band, this phase can be arbitrarily chosen using a gauge transformation, $(c_k, c^\dagger_{-k}) \ra (e^{i\theta/2}c_k, e^{-i\theta/2}c^\dagger_{-k})$, $\Delta_l \ra e^{i\theta}\Delta_l$. The edge dispersion and the edge current are unaffected by the transformation. However, for two coupled bands, the phases of different bands cannot be individually gauge-transformed. Hence the relative phase between the quasiparticle wave functions (\ref{eq:wavefunction}) bears nontrivial physical significance. To see how interband mixing affects the chiral edge modes, we evaluate the hybridization between two edge states associated with the two bands, $\phi_{1,k_y}$ and $\phi_{2,k_y^\prime}$,
\begin{equation}
\ep_{12}(k_y,k_y^\prime) =  \langle \phi_{1,k_y} |H^\prime_{12}| \phi_{2,k_y^\prime} \rangle  \,,
\label{eq:MatrixElement}
\end{equation}
where $H^\prime_{12}$ denotes the mixing term (\ref{eq:mixing}) in Nambu space: $H^\prime_{12} = \tau_3 g(x)$. Here $\tau_3$ is the third component of the Pauli matrix and $g(x)$ is an unimportant function characteristic of the spatial dependence of the interband mixing at the edge. For finite $\ep_{12}$ the two modes hybridize and level-split. Clearly, for time-reversal invariant interband mixing, i.e., for real $t_{12}$,
\begin{equation}
\ep_{12}(k_y, k_y^\prime) \propto \left\{
        \begin{array}{ll}
            0 & ~~\text{sgn}[\Delta_1] = \text{sgn}[\Delta_2], \\
            |t_{12}|\delta_{k_y,k_y^\prime} & ~~\text{sgn}[\Delta_1] = -\text{sgn}[\Delta_2] .
        \end{array}
    \right.
\end{equation}
The splitting is thus readily understood.

The piling-up of the subgap edge density of states at higher energy can be straightforwardly analyzed in a similar fashion, except that in this case the hybridization with the (bulk) continuum states (which situate closer in energy with respect to those edge modes) becomes influential in both phase configurations. Note the bulk states generally have unequal electron and hole components, hence the hybridization typically does not vanish at first order in $t_{12}$ in the $[++]$ model.

\subsection{Effective field theory}
\label{subsec:EFT}
In this section we provide a phenomenological understanding of the interference-induced variations in the behavior of the order parameter and the edge current.

At an ideal edge the $\Delta_{x}$ components must fall to zero, as they change sign under a reflection about $y$. By contrast, the $\Delta_{y}$ components typically remain finite. The interband mixing induces free-energy terms $\propto K|\Delta_{1x(y)}-\Delta_{2x(y)}|^2$ localized at the boundary, which favor a vanishing (phase) difference between $\Delta_1$ and $\Delta_2$. Here the coefficient $K$ parametrizes the strength of the mixing. These terms resemble the gradient energy cost in the usual Ginzburg-Landau theory. Consequently, in the $[+-]$ configuration and when the mixing is strong, one of the $\Delta_{y}$'s tends to change sign to minimize the induced free energy. This leads to a simultaneous response in the $\Delta_x$ component of the same band, which can be understood from higher-order terms, such as $-|\Delta_x|^2|\partial_x\Delta_y|^2$, which supports larger $\Delta_x$ when $\Delta_y$ varies\cite{Matsumoto:99}.

It is known that the edge current is related to the spatial gradients of the order parameter components\cite{Sigrist:91,Furusaki:01,Huang:14,Huang:15}, and at lowest order the relation can be described as follows \cite{Huang:15}:
\begin{equation}
j_y \sim \sum_{l=1,2} \gamma_l \partial_x  \left [ \frac{ \Delta_{ly}}{\Delta_{ly0}} - \frac{ \Delta_{lx}}{\Delta_{lx0}}   \right ] \,,
\label{eq:GL}
\end{equation}
where $\Delta_{lx0}$ and $\Delta_{ly0}$ are the bulk values of the chiral components; $\gamma_l$ is a phenomenological parameter characteristic of the band and gap structure of the $l$ band, given explicitly by \cite{Huang:15} $\sum_{\bs k} v_{lx}v_{ly}f_{l,\bs k}g_{l,\bs k} |\Delta_{lx0} \Delta_{ly0}|/(2E_{l,\bs k}^3)$, where $E_{l,\bs k}=\sqrt{\varepsilon_{l,\bs k}^2 + |\Delta_{l,\bs k}|^2}$ and $v_{lx(y)}$ are components of the Fermi velocity. It is then easy to see how variations in the spatial profile of the order parameters go hand in hand with those in the edge current. As can be seen in Fig. \ref{fig:curr}, this phenomenological expression indeed roughly captures the behavior of the edge current distribution.


\section{Edge tunneling conductance}
\label{sec:BTK}
In this section, we examine the consequence of interband interference on the edge tunneling conductance within the standard classical BTK theory \cite{Blonder:82}. We base the following calculations on an important similar study by Golubov {\it et al.}\cite{Golubov:09} formulated in the context of multiband $s$-wave models\cite{Golubov:09}. Single-band analyses can be found in, e.g., Ref \onlinecite{Yamashiro:98}.

\subsection{BTK calculations}
\label{subsec:BTKintro}
We consider a junction formed by a one-band normal metal (N) and a two-band superconductor (S), and for simplicity take the bands on both sides of the junction to have the same band structure and Fermi surface. The wavefunction on the N-side reads,
\begin{equation}
\Psi_N(\bs r) =
\left[
\begin{pmatrix}
1 \\ 0
\end{pmatrix}
+a\begin{pmatrix}
0 \\ 1
\end{pmatrix}
\right]
e^{i(k_x x+k_y y)}
+b\begin{pmatrix}
1 \\ 0
\end{pmatrix}e^{i(-k_x x+k_y y)},
\end{equation}
and on the S-side,
\begin{eqnarray}
&& \Psi_S (\bs r) = c\left[  \begin{pmatrix} u_{1,\bs k} \\ h_{1,\bs k} v_{1,\bs k} \end{pmatrix}
+ \alpha  \begin{pmatrix} u_{2,\bs k} \\ e^{i\varphi}h_{2,\bs k} v_{2,\bs k} \end{pmatrix} \right]e^{i(k_x x +k_y y)}  \nonumber \\
&&+
d\left[  \begin{pmatrix} v_{1,\bs k^\prime} \\ h_{1,\bs k^\prime} u_{1,\bs k^\prime} \end{pmatrix}
+ \alpha  \begin{pmatrix} v_{2,\bs k^\prime} \\ e^{i\varphi}h_{2,\bs k^\prime} u_{2,\bs k^\prime} \end{pmatrix} \right]e^{i(-k_x x + k_y y)}\,, \nonumber \\
&&
\end{eqnarray}
where ${\bs k}=(k_x,k_y)$ and ${\bs k}^\prime=(-k_x,k_y)$ represent, respectively, the incident Fermi wavevector in the metal and the specularly reflected wave vector, $h_{l,\bs k} = \Delta_{l,\bs k}/|\Delta_{l,\bs k}|$, $u_{l,\bs k} (v_{l,\bs k}) = \sqrt{ \frac 12 \left[ 1 +(-) \sqrt{E^2-|\Delta_{l,\bs k}|^2}/E \right]}$, in which $E$ is the quasiparticle energy, $\varphi$ denotes the phase difference between the two bands, and $\alpha$ is a coefficient which determines the relative probability for a normal metal electron to tunnel into the two superconducting bands. Hence $\alpha = 0$ or $ \infty$ returns a one-band NS junction. Since its value depends on the property of the two superconducting bands \cite{Golubov:09}, we take it as a tuning parameter. Noteworthily, although $\alpha$ thus defined does not rely on the presence of edge disorder and is therefore not in one-to-one correspondence with $t_{12}$ introduced in BdG, it does reflect a certain level of interband mixing. Hence a similar flavor of interband interference is also expected here, as we shall see below.

The solutions are determined by the following boundary conditions at the NS interface,
\begin{eqnarray}
\label{boundary}
&&\Psi_N(x=0,y)=\Psi_S(x=0,y), \notag \\
&&\frac{\partial\Psi_S}{\partial x}\biggl\rvert_{x=0^+} - \frac{\partial\Psi_N}{\partial x}\biggl\rvert_{x=0^-}=\frac{2m H}{\hbar^2}\Psi_S|_{x=0^-},
\end{eqnarray}
where $H$ characterizes the strength of the $k$-independent potential barrier. Define a dimensionless parameter $Z \equiv \frac{mH}{\hbar^2 k_{x}}$; since $Z \geq \frac{mH}{\hbar^2k_F}$, it can be verified that in the tunneling limit, i.e., for sufficiently large $H$, the $k_x$ dependence of the conductance is insignificant. Hence we take a fixed value $Z=10$ in the calculations. More details are given in the Appendix.

\begin{figure}
\subfigure{ \includegraphics[width=4cm]{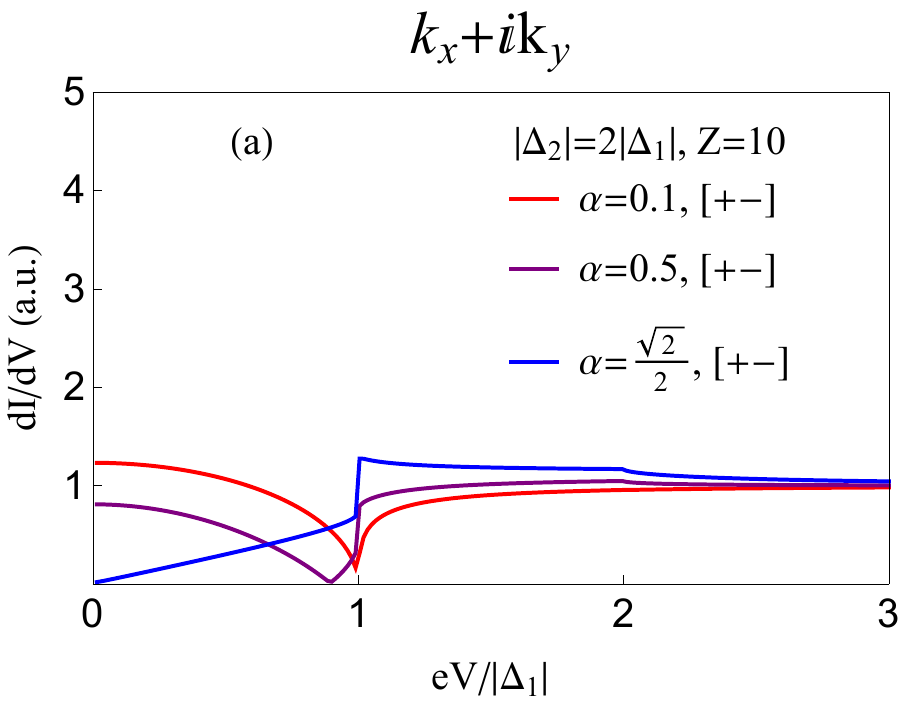} }
\subfigure{ \includegraphics[width=4cm]{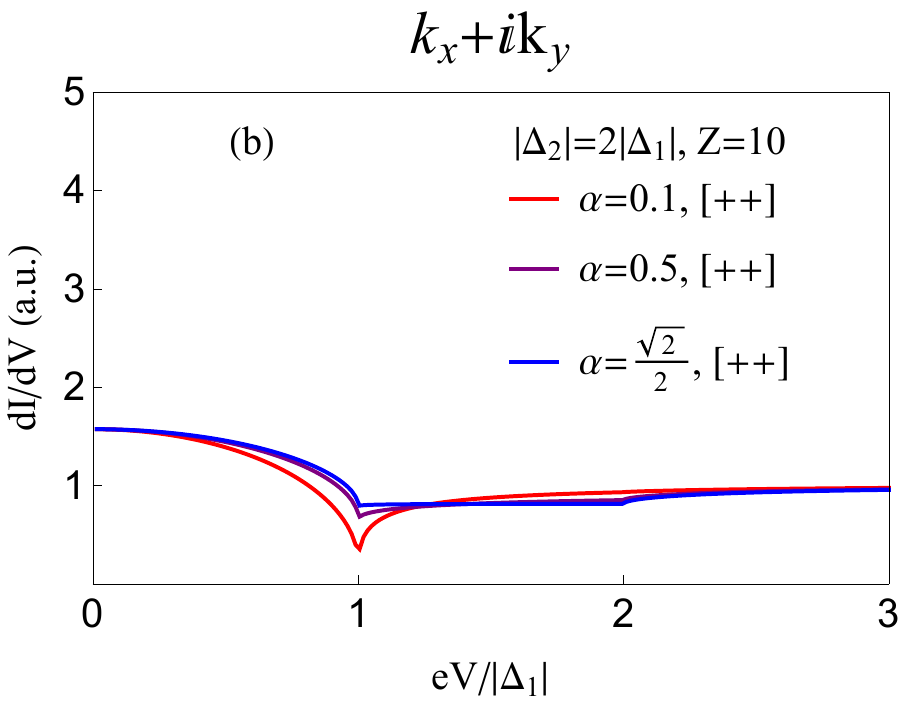} }
\caption{(Color online) Tunneling conductance spectra of two-band chiral $p$-wave models at a NS junction. For calculational convenience the Fermi surfaces on both sides of the junction are taken to be circular and equal-size with $k_F=1$. Their gap functions assume the same form, $\Delta_l(k_x + i k_y)$, with the amplitudes satisfying the relation $|\Delta_1|=2|\Delta_2|$.}
\label{fig:chiralP}
\end{figure}

\begin{figure}
\subfigure{ \includegraphics[width=4cm]{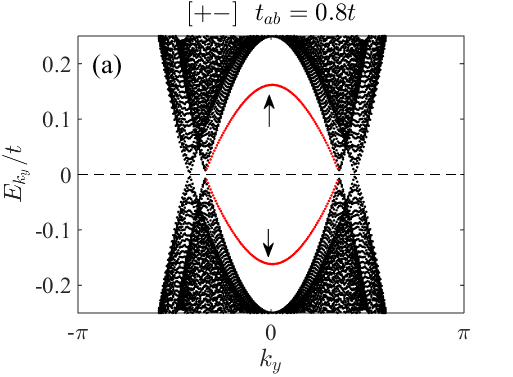} }
\subfigure{ \includegraphics[width=4cm]{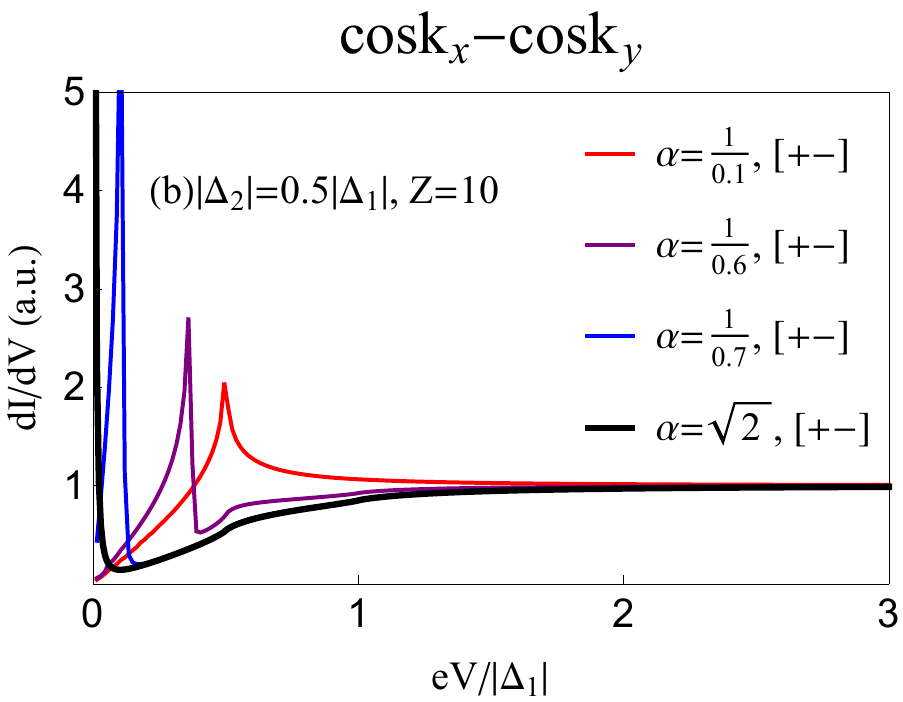} }
\caption{(Color online) (a) BdG spectrum of a two-band $d_{x^2-y^2}$-wave model with the $[+-]$ phase configuration with interband mixing at one of the edges. Part of the dispersive edge modes is depicted in red (as indicated by the arrows). The calculation is performed using the same band structure as given in Fig. \ref{fig:disp1}, and the gap functions of the two bands are: $\Delta_{1/2,\bs k} \propto \cos k_x - \cos k_y$, with gap amplitudes $0.2t$ and $0.25t$, respectively.  (b) Conductance spectra of a two-band $d$-wave model with the $[+-]$ configuration. The gap functions are given by $\Delta_l(\cos k_x - \cos k_y)$. Note the zero-bias peak (black curve), which does not exist at the $x$ or $y$ edges of a single-band $d_{x^2-y^2}$ superconductor.}
\label{fig:Dwave}
\end{figure}

Figure \ref{fig:chiralP} presents the conductance spectra of a simple two-band chiral $p$-wave model. Consistent with the BdG results, the influence of the interband interference is selective between the two phase configurations. In the $[++]$ model, the low-bias portion of the spectra hardly varies, irrespective of the presence of noticeable interband mixing [Fig \ref{fig:chiralP} (b)]. By contrast, in the $[+-]$ model the spectra are vastly modified and in particular, the zero-bias conductance characteristic of the single-band spectra has almost disappeared at the same level of $\alpha$ \cite{footnote}. These closely resemble the observations in our BdG calculations, where the low-energy DOS is depleted in the $[+-]$ model but not in the $[++]$ model. In addition, the noticeable enhancement of the conductance near the lower gap in the $[+-]$, as well as the moderate enhancement in the $[++]$ model, can be related to the piling up of a high-energy edge DOS seen in BdG. As shown in Fig. \ref{fig:chiralPapp}, lattice models can even see a peak developing at the lower gap energy, but this is not typical.

\subsection{Other pairing symmetries}
\label{subsec:otherPairing}
As a further remark, many other factors, such as the barrier potential variation across the tunneling junction, can seriously modify the conductance spectra. It was argued \cite{Kashiwaya:11} that the sensitivity of the spectra to microscopic details, combined with the existence of subgap conductance peaks \cite{Laube:00,Kashiwaya:11,Ying:12,Yada:14}, provides strong support for chiral $p$-wave pairing in \SRO. Interestingly, the existence of the subgap Andreev states and the rich variety of spectral shapes are, in fact, common to other pairing symmetries, pending a multiband nature and a nontrivial phase configuration. These have been shown in a number of theoretical works on the $s_{\pm}$-wave models formulated for, e.g. the multiband iron-based superconductors \cite{Golubov:09,Ghaemi:09,Bobkov:11}. The same conclusions apply to other multiband systems whose single-band counterparts do not, in principle, support subgap bound states, such as the $d_{x^2-y^2}$-wave pairing exemplified in Fig. \ref{fig:Dwave}. As one can see, the conductance peak shifts as the interband mixing varies, and for certain $\alpha$ a zero-bias-like peak emerges which can hardly be distinguished from that of a single-band $d_{xy}$ superconductor \cite{Tanaka:95,Hu:94}[Fig. \ref{fig:Dwave} (b)]. These strong interband interference effects hold potential relevance for, e.g., the heavy-fermion superconductor CeCoIn$_5$ \cite{Park:08}, and similarly for \SRO~if it supports multiband even-parity pairing \cite{Zhang:17}.


\section{Summary}
\label{sec:summary}
In this work, we studied how interband interference peculiar to multiband systems may operate to qualitatively change the physics at the edges of a two-band chiral $p$-wave superconductor. We showed that, in systems possessing nonvanishing phase differences between the bands, interband interference can strongly influence the edge spectrum, which then results in observable variations in the edge current distribution and the edge tunneling conductance. These results are discussed in conjunction with \SRO~-- a candidate chiral $p$-wave superconductor with multiband character. Without serious modification, the conclusions can be directly generalized to other putative multiband chiral superconductors, such as UPt$_3$ \cite{Joynt:03} and SrPtAs \cite{Fischer:14}. In passing, we also pointed out (on the basis of the present and previous works) that the formation of subgap edge states and their sensitivity to the interband mixing are universal in multiband superconductors with other pairing symmetries.

\section{Acknowledgement}
\label{sec:ack}
We would like to thank Y. Tanaka for a helpful communication. This project is supported by Grants No. NKRDPC-2017YFA0206203, No. NSFC-11574404, NSFC-11275279, No.NSFG-2015A030313176, the Special Program for Applied Research on Super Computation of the NSFC-Guangdong Joint Fund (the second phase), the National Supercomputer Center in Guangzhou, and the Leading Talent Program of Guangdong Special Projects (D.-X.Y. and J.-L.Z.), as well as the C. N. Yang Junior Fellowship at Tsinghua University (W.H.).

\section{Appendix}
\begin{figure}
\includegraphics[width=5cm]{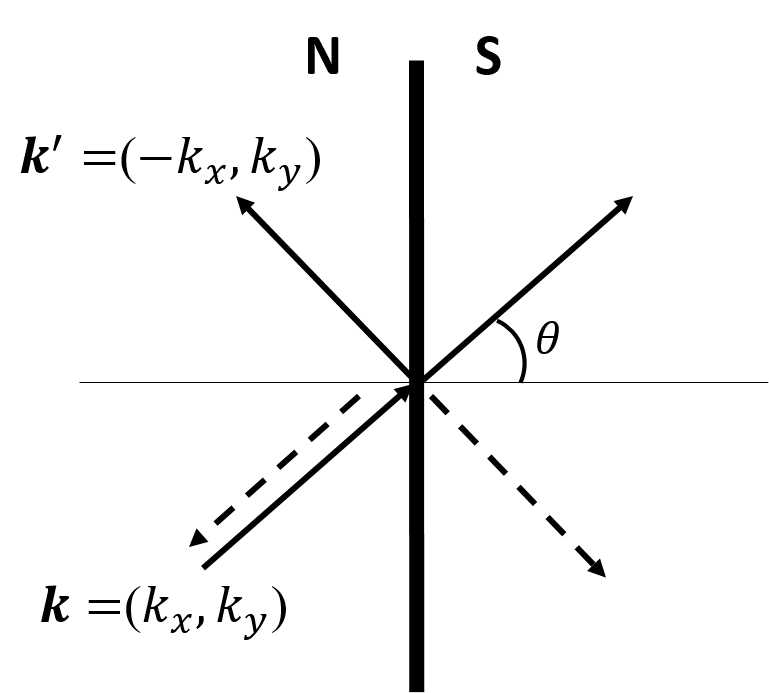}
\caption{Schematic diagram of a normal-superconductor (NS) junction. The trajectories of the electron- and hole-like quasiparticles are designated by solid and dashed lines, respectively. The band indicies are implicit, and we have simplified the diagram by assuming that the Fermi surface of the two bands are the same. So the wavevectors of the two bands on the S side coincide.}
\label{fig:Andreev}
\end{figure}

To simulate the tunneling conductance between a normal metal and a two-band superconductor, we adopt the approach developed by Golubov {\it et al.} \cite{Golubov:09}, which is formulated for $s$-wave models. Consider a two-dimensional normal metal-superconductor (NS) junction (Fig \ref{fig:Andreev}), the overall wavefunction can be written as,
\begin{equation}
\Psi(\mathbf{r}) = \Psi_N(\mathbf{r}) \Theta(-x)+\Psi_S(\mathbf{r})\Theta(x),
\end{equation}
where on the normal side,
\begin{equation}
\Psi_N =
\left[
\begin{pmatrix}
1 \\ 0
\end{pmatrix}
+a\begin{pmatrix}
0 \\ 1
\end{pmatrix}
\right]
e^{i(k_x x+k_y y)}
+b\begin{pmatrix}
1 \\ 0
\end{pmatrix}e^{i(-k_x x+k_y y)},
\end{equation}
${\bs k}=(k_x,k_y)$ is a wave vector near the Fermi surface, and on the superconductor side with two bands,
\begin{eqnarray}
\Psi_S &=& c
\begin{pmatrix}
u_{1,(q_x,q_y)} \\ v_{1,(q_x,q_y)} \frac{\Delta_{1,(q_x,q_y)}} {|\Delta_{1,(q_x,q_y)}|}
\end{pmatrix}  e^{i(q_x x+q_y y)}  \nonumber \\
&+& \alpha c
\begin{pmatrix}
u_{2,(p_x,p_y)} \\ e^{i \varphi} v_{2,(p_x,p_y)} \frac{\Delta_{2,(p_x,p_y)}} {|\Delta_{2,(p_x,p_y)}|}
\end{pmatrix}  e^{i(p_x x+p_y y)}  \nonumber \\
&+&d
\begin{pmatrix}
v_{1,(-q_x,q_y)} \\ u_{1,(-q_x,q_y)} \frac{\Delta_{1,(-q_x,q_y)}} {|\Delta_{1,(-q_x,q_y)}|}
\end{pmatrix}  e^{i(-q_x x+q_y y)} \nonumber \\
&+& \alpha d
\begin{pmatrix}
v_{2,(-p_x,p_y)} \\ e^{i \varphi} u_{2,(-p_x,p_y)} \frac{\Delta_{2,(-p_x,p_y)}} {|\Delta_{2,(-p_x,p_y)}|}
\end{pmatrix}  e^{i(-p_x x+p_y y)} \,, \nonumber \\
&&
\end{eqnarray}
where $u_{1(2),(k_x,k_y)}  = \sqrt{ \frac 12 (1 + \frac{\sqrt{E^2-|\Delta_{1(2),(k_x,k_y)}|^2}}{E}) } $, $v_{1(2),(k_x,k_y)}  = \sqrt{ \frac 12 (1 - \frac{\sqrt{E^2-|\Delta_{1(2),(k_x,k_y)}|^2}}{E}) } $. Note that for both $\Psi_N$ and $\Psi_S$, the momentum along $y$ is conserved, as can be seen by the common factor $e^{ik_y y}$ in both expressions. In $\Psi_S$ the coefficient $\alpha$ is a tunable parameter which determines the relative tunneling probabilities onto the two superconducting bands. When $\alpha=0$ and $\alpha \gg 0$, our model returns to the one-band case.

\begin{figure}
\subfigure{ \includegraphics[width=4cm]{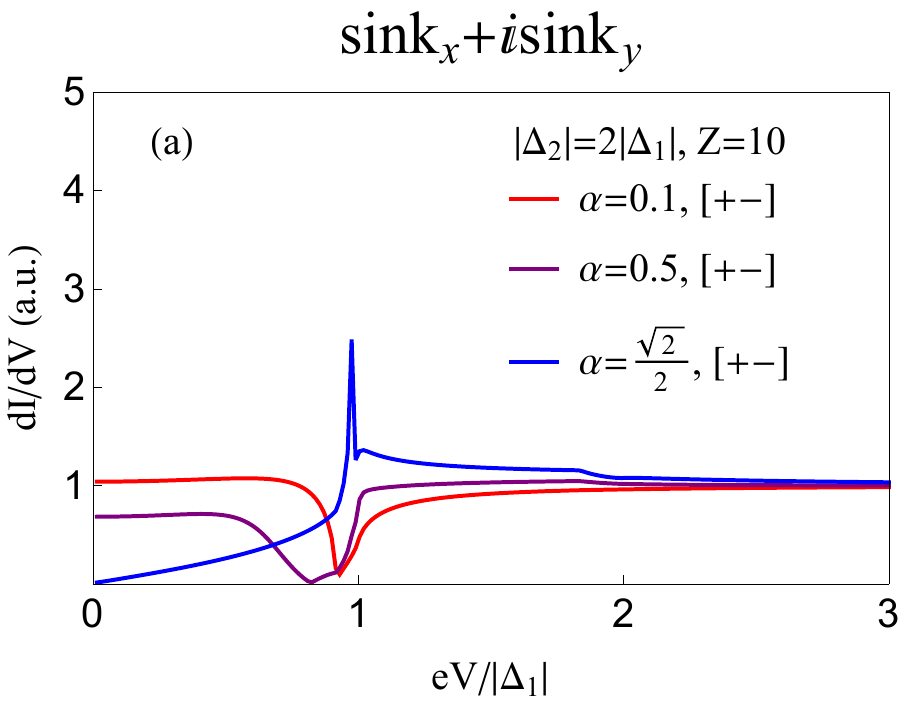} }
\subfigure{ \includegraphics[width=4cm]{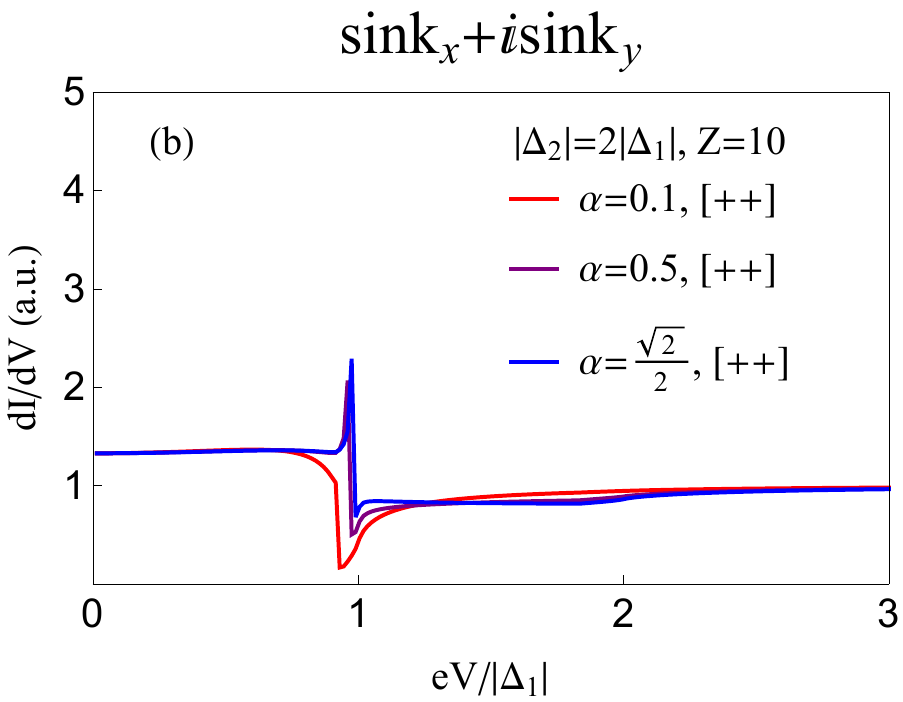} }
\subfigure{ \includegraphics[width=4cm]{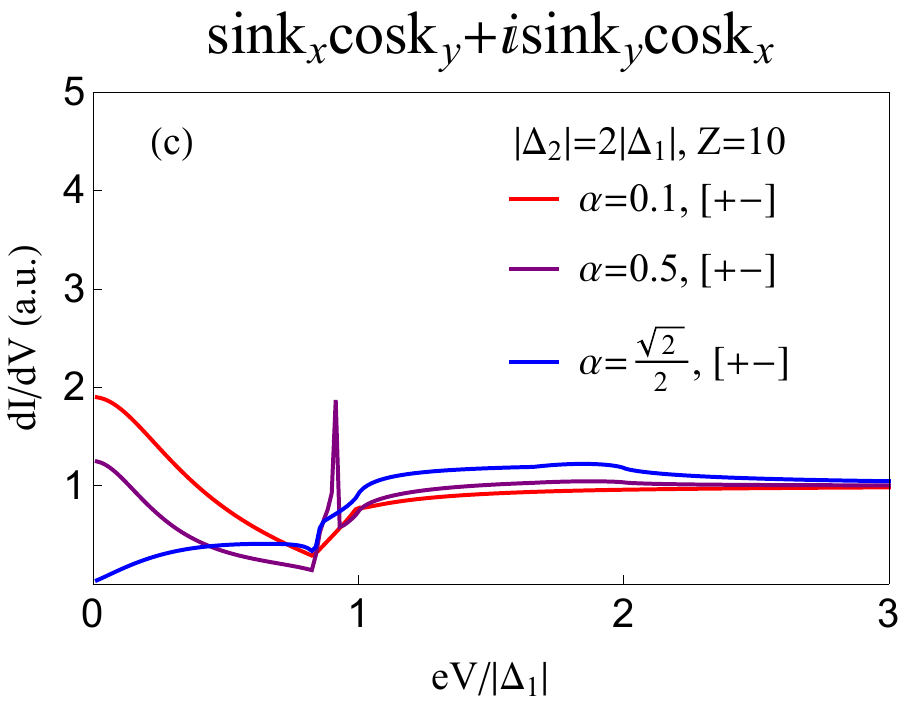} }
\subfigure{ \includegraphics[width=4cm]{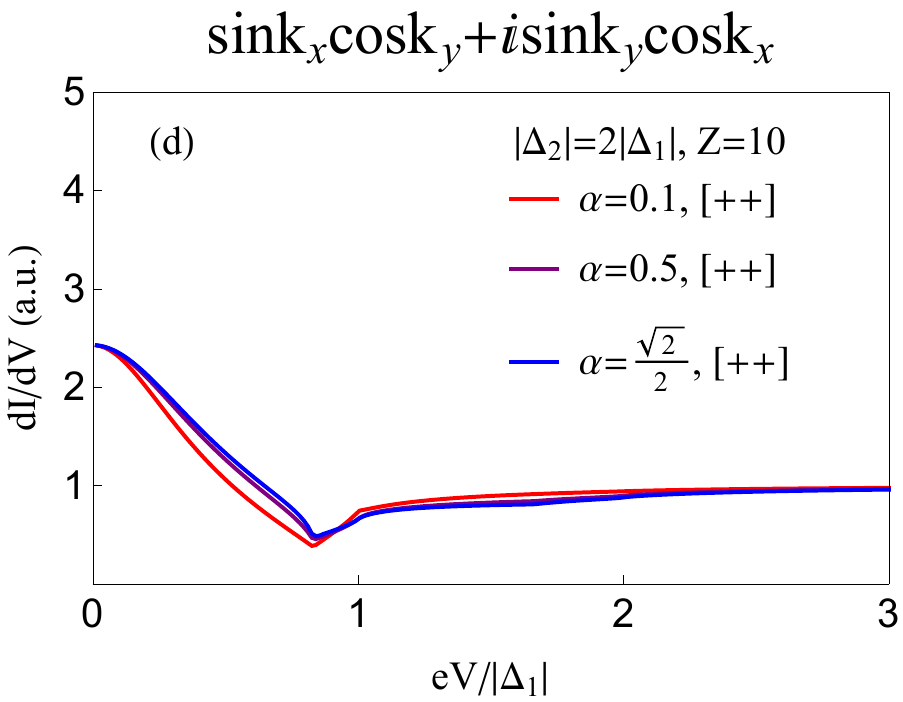} }
\caption{(Color online) Tunneling conductance spectra of two-band chiral $p$-wave models with different phase configurations at a NS junction. For calculational convenience the Fermi surfaces on both sides of the junction are taken to be circular and equal size with $k_F=1$. Their gap functions assume the same form: (a) and (b) $\Delta_l(\sin k_x + i \sin k_y)$ and (c) and (d) $\Delta_l(\sin k_x\cos k_y + i\cos k_x \sin k_y)$. The amplitudes of the two gaps have the relation $|\Delta_1|=2|\Delta_2|$ in all calculations shown. }
\label{fig:chiralPapp}
\end{figure}

\begin{figure}
\subfigure{ \includegraphics[width=4.cm]{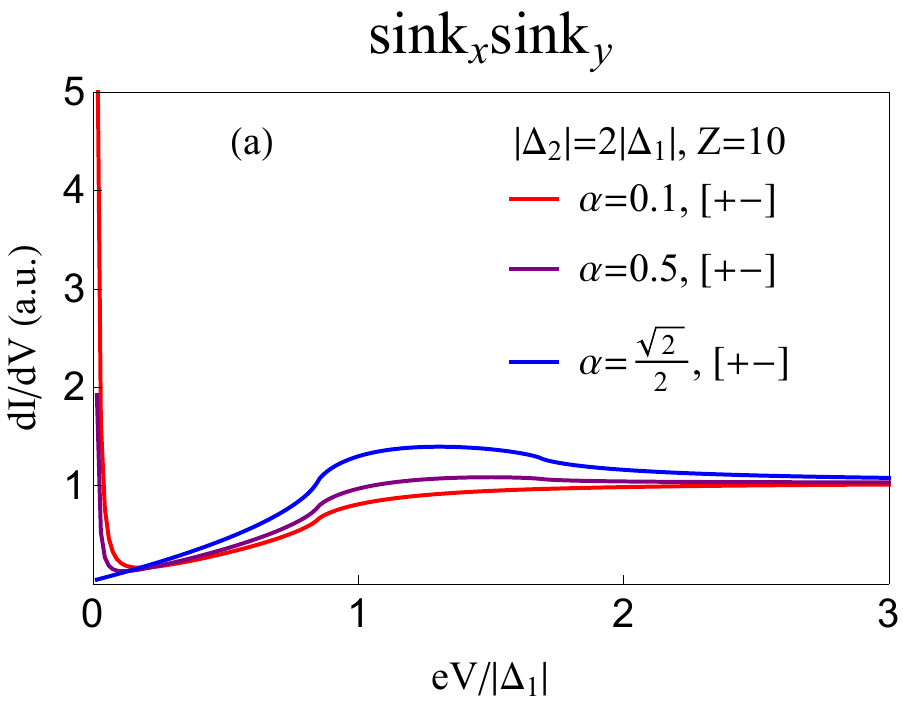}}
\subfigure{ \includegraphics[width=4.cm]{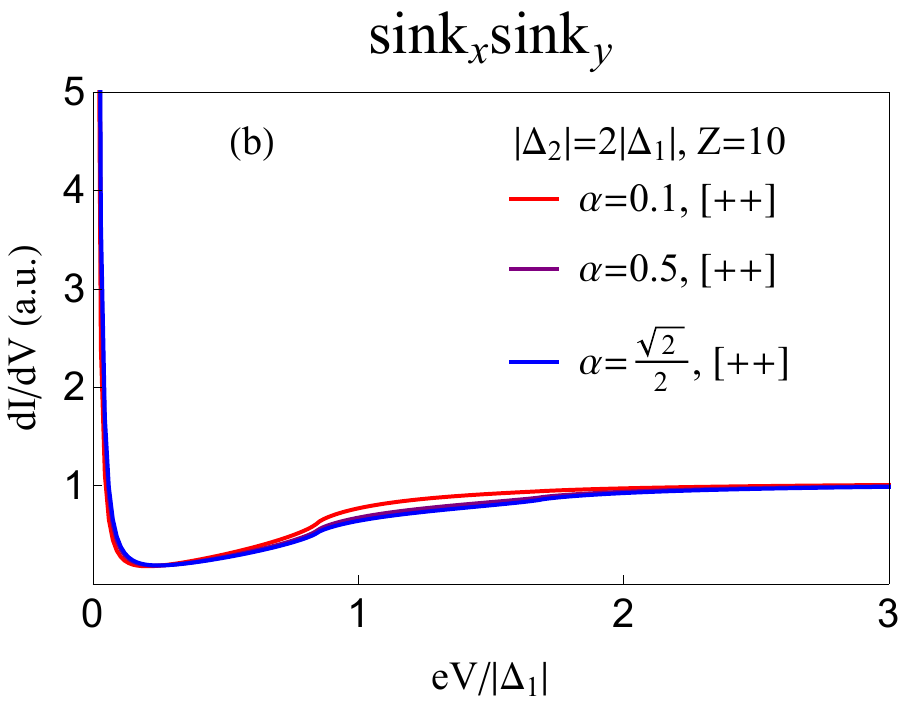}}
\caption{(Color online) Tunneling conductance spectra at a NS junction of a two-band $d_{xy}$-wave model with (a) $[+-]$ and (b) $[++]$ phase configurations. The Fermi surfaces of the metal and the two bands of the superconductor are taken to be circular and equal size with $k_F=1$. Their gap functions are given by $\Delta_{l,\bs k} =\Delta_{l} \sin k_x \sin k_y$ with $|\Delta_1|=2|\Delta_2|$. }
\label{fig:Dxy}
\end{figure}

The boundary conditions at the interface of the junction are
\begin{eqnarray}
\label{boundary}
&&\Psi_N|_{x=0^-}=\Psi_S|_{x=0^+}, \notag \\
&&\frac{\partial\Psi_S}{\partial x} \biggl\rvert_{x=0^+} - \frac{\partial\Psi_N}{\partial x}\biggl\rvert_{x=0^-}=\frac{2m H}{\hbar^2}\Psi_S \rvert_{x=0^+},
\end{eqnarray}
where $H$ characterizes the strength of the $k$-independent potential barrier. The tunneling limit corresponds to an $H$ which is large compared with $E_F$. Without loss of any essential physics, we assumed that the normal side and the superconducting side have the same Fermi surface, i.e., $(k_x,k_y)=(q_x,q_y)=(p_x,p_y)$. Defining $Z \equiv \frac{mH}{\hbar^2 k_{x}}$, which is always larger than $\frac{mH}{\hbar^2k_F}$, it can be easily shown that the spectrum hardly changes beyond a sufficiently large $H$. On this account we choose a $k$-independent parameter $Z=10$. The coefficients $a$, $b$, $c$, and $d$ satisfy the following relations,
\begin{eqnarray}
&&\gamma_0 a=\gamma_3 \gamma_4 \, \nonumber \\
&&\gamma_0 b=(Z^2-iZ)(\gamma_2 \gamma_3-\gamma_1 \gamma_4) \,, \nonumber\\
&&\gamma_0 c=(1-iZ)\gamma_4 \,, \nonumber\\
&&\gamma_0 d=iZ\gamma_3 \,,
\end{eqnarray}
where
\begin{eqnarray}
\gamma_1&=&u_{1,(q_x,q_y)}+\alpha u_{2,(p_x,p_y)} \,, \nonumber \\
\gamma_2&=&v_{1,(-q_x,q_y)}+\alpha v_{2,(-p_x,p_y)} \,, \nonumber \\
\gamma_3&=&v_{1,(q_x,q_y)} \frac{\Delta_{1,(q_x,q_y)}} {|\Delta_{1,(q_x,q_y)}|}+\alpha e^{i \varphi} v_{2,(p_x,p_y)} \frac{\Delta_{2,(p_x,p_y)}} {|\Delta_{2,(p_x,p_y)}|} \,, \nonumber \\
\gamma_4&=&u_{1,(-q_x,q_y)} \frac{\Delta_{1,(-q_x,q_y)}} {|\Delta_{1,(-q_x,q_y)}|}+\alpha e^{i \varphi} u_{2,(-p_x,p_y)} \frac{\Delta_{2,(-p_x,p_y)}} {|\Delta_{2,(-p_x,p_y)}|} \,, \nonumber\\
\gamma_0 &=&(1+Z^2)\gamma_1 \gamma_4-Z^2\gamma_2 \gamma_3.
\end{eqnarray}
Notice that in the above expressions we have not yet used the assumption ${\bs k}={\bs q}={\bs p}$ for the purpose of generality. The remaining calculations nevertheless use this relation throughout. The differential conductance of the NS junction is defined as,
\begin{equation}
 \sigma_S(\theta)=1+|a|^2-|b|^2,
\label{btk}
\end{equation}
where $\theta$ is the injection angle; thus we have $\bs{k}=(k_x,k_y)=k(\cos\theta,\sin\theta)$, and $|a|^2$ and $|b|^2$ are the Andreev reflection and normal refection probabilities, respectively. In a two-dimensional model, the total tunneling conductance is the integration of $\sigma_S$ over the injection angle $\theta$,
\begin{equation}
  \frac{dI}{dV}=\frac{\int_{-\frac{\pi}{2}}^{\frac{\pi}{2}} \sigma_S(\theta)\cos(\theta)d\theta} {\int_{-\frac{\pi}{2}}^{\frac{\pi}{2}} \sigma_N(\theta)\cos(\theta)d\theta},
\end{equation}
where $\sigma_N$ is the conductance when superconductivity on the S side vanishes.

To complement the tunneling spectra shown in the main text, Figs. \ref{fig:chiralPapp} and \ref{fig:Dxy} display, respectively, the tunneling conductance spectra of some lattice models of two-band chiral $p$-wave and $d_{xy}$-wave superconductors. As is obvious, when the two band gaps are opposite in sign, the zero-bias features expected for a single-band model can be entirely destroyed by the interband interference in the presence of finite phase difference between the two bands.


\begin{thebibliography}{99}
\bibitem{footnote} Here, the value of $\alpha$ at which the interband mixing has the strongest effect is not universal. It depends on the ratio of the two gaps \cite{Golubov:09}.
\end{thebibliography}
\end{document}